\documentstyle[12pt,fleqn]{article}
\def\mbi#1{\mbox{\boldmath$#1$}}
\def\kp{\mbi{k} \cdot \mbi{p}}

\def\la{\langle}
\def\ra{\rangle}
\def\beeq{\begin{equation}}
\def\eneq{\end{equation}}
\def\beeqa{\begin{eqnarray}}
\def\eneqa{\end{eqnarray}}

\addtocounter{section}{-1}
\begin{document}

\begin{center}

\vspace{2cm}

{\large {\bf {Magnetic impurity effects
in metallic carbon nanotubes:\\
local non-Fermi liquid theory} } }

\vspace{1cm}

{\rm Kikuo Harigaya\footnote[1]{E-mail address: 
\verb+harigaya@etl.go.jp+; URL: 
\verb+http://www.etl.go.jp/+\~{}\verb+harigaya/+}}

\vspace{1cm}

{\sl Physical Science Division,
Electrotechnical Laboratory,\\ 
Umezono 1-1-4, Tsukuba 305-8568, 
Japan}\footnote[2]{Corresponding address}\\
{\sl National Institute of Materials and Chemical Research,\\ 
Higashi 1-1, Tsukuba 305-8565, Japan}\\
{\sl Kanazawa Institute of Technology,\\
Ohgigaoka 7-1, Nonoichi 921-8501, Japan}

\vspace{1cm}
(Received~~~~~~~~~~~~~~~~~~~~~~~~~~~~~~~~~~~)
\end{center}

\vspace{1cm}

\noindent
{\bf Abstract}\\
Magnetic impurity effects on metallic carbon nanotubes
are studied theoretically.  The resolvent method for the
multi channel Kondo effect is applied to the band structure
of the $\kp$ perturbation hamiltonian in the limit of the
infinite onsite repulsion at the impurity site.  We discuss 
the local non-Fermi liquid behavior at temperatures lower 
than the Kondo temperature $T_{\rm K}$.  The density of states
of localized electron has a singularity $\sim |\omega|^{1/2}$
which gives rise to a pseudo gap at the Kondo resonance
in low temperatures.  The temperature dependence of the 
electronic resistivity is predicted as $T^{1/2}$, and the
imaginary part of dynamical susceptibilities has the
$|\omega|^{1/2}$ dependence.  Possible experimental 
observations are discussed.

\vspace{1cm}

\pagebreak

\section{Introduction}

Recently, carbon nanotubes with cylindrical graphite 
structures have been intensively investigated.  Many interesting 
experimental as well as theoretical researches have been 
performed (see reviews [1,2] for example), and the fundamental
metallic and semiconducting behaviors of single wall nanotubes
predicted by theories [3-8] have been clarified in tunneling
spectroscopy experiments [9,10].

In the magnetoresistance study of metallic nanotubes [11], 
disorder gives rise to the positive differential resistance 
at low temperatures due to the weak localization effects [12].  
Some samples are known to show the negative resistance 
around the weak gate voltage region, which is interpreted 
possibly due to the presence of the Kondo resonance as the 
result of the magnetic impurity effect.  Therefore, a 
magnetic impurity might change the electronic structures of 
metallic carbon nanotubes.

The second candidate of the magnetic impurity is the fluorine 
(F) adatoms on the nanotubes.  The recent experiment [13,14] 
of the nanographite doped with F shows appearance of electronic
spin during the fluorine concentration, $[{\rm F}]/[{\rm C}] = 0 \sim 1.2$.
The computer simulation of F-doping effects on nanographite [15]
supports the presence of a local spin at the fluorine atom
bonded to the carbon atom.

Thirdly, it is known that pristine carbon nanotubes are 
found to coexist with amorphous carbon soots [1,2].  Their effects
are not of prime interests usually.  However, we are aware
that amorphous metals are modeled well by the two level
system [16,17].  The low temperature behavior of resistivity
in the two level system is similar to that of the Kondo model 
of the magnetic impurity [18,19,20].  So, the metallic
nanotube with amorphous carbon soots may show behaviors like 
effects of magnetic impurities at low temperatures.  This 
viewpoint is very interesting as a possible realization
of the results of low temperature properties of the two
level system, or equivalently, magnetic impurity effects.

In this paper, we will study effects of a magnetic
impurity in metallic carbon nanotubes at low temperatures.
The nonmagnetic impurity effects [21,22] have been studied
by the present author, too.  However, magnetic impurity effects
are attractive also, when we look at our experience of the 
research of $f$-electron systems [23].  There are two channels
of electronic states at the Fermi energy in metallic carbon
nanotubes.  As known in the magnetic systems, such as, heavy
fermion systems [24], the non-Fermi liquid behaviors, i.e.,
the singular density of states and the power law temperature
dependence of the electric resistivity, have been observed 
experimentally and explained theoretically by using the Kondo 
model or the Anderson model with multi channel scatterings.
The similar effects can occur in the carbon nanotubes when the
presence of the two scattering channels plays an important role.

We will use the $\kp$ method [25,26] for the electronic
states of carbon nanotubes.  The method well describes 
electronic states around the Fermi energy.  The valence
and conduction band states have the linear dispersion
at the Fermi energy in metallic nanotubes.  It is 
assumed that there is one magnetic impurity at a
carbon site.  The magnetic impurity is modeled by the
Anderson model where there is one localized electronic
state and the strong onsite repulsion is assumed.
For the treatment of the Kondo effect, we use the
resolvent method [27,28] to discuss the infinite
repulsion case.  This is sufficient for the discussion
of the low temperature and low energy behaviors.
We will solve spectral functions of the resolvents,
and derive analytic formulas of electronic density
of states, resistivity, and dynamical susceptibilities.
We will discuss the local non-Fermi liquid behavior 
which could be observed at low temperatures.  The 
density of states of localized electron has a singularity 
$\sim |\omega|^{1/2}$.  This singular behavior gives 
rise to a pseudo gap at the Kondo resonance in low 
temperatures.  The temperature dependence of the 
electronic resistivity is predicted as $T^{1/2}$, 
and the imaginary part of dynamical susceptibilities 
has the $|\omega|^{1/2}$ dependence.  Possible experimental
observations are discussed.

This paper is organized as follows.  In the next section,
we introduce the model and explain theoretical formulations.
In Sec. III, we report the low temperature solution
of resolvents.  In Sec. IV, electronic density of
states is derived.  In Secs. V and VI, resistivity
and dynamical susceptibilities are explained.
Section VII is devoted to discussion, and the
paper is closed with summary in Sec. VIII.

\section{Model}

We will study the metallic carbon nanotubes with one
Anderson impurity at the A or B sublattice site.  In the
total hamiltonian,
\beeq
H = H_{\rm tube} + H_{\rm imp},
\eneq
$H_{\rm tube}$ is the electronic states of the carbon
nanotubes, and the model based on the $\kp$ approximation [25,26]
represents electronic systems on the continuum medium.
The second term $H_{\rm imp}$ is the Anderson impurity
where the onsite Coulomb interaction strength is taken
as infinite in the course of calculation, and the resolvent 
formalism [27,28] is introduced in order to take into account 
of the infinite repulsion.

The hamiltonian by the $\kp$ approximation [25,26] in the
secondly quantized representation has the following
form:
\beeq
H_{\rm tube} = \sum_{\mbi{k},\sigma} \Psi_{\mbi{k},\sigma}^\dagger
E_{\mbi{k}} \Psi_{\mbi{k},\sigma},
\eneq
where $E_{\mbi{k}}$ is an energy matrix:
\beeq
E_{\mbi{k}} =
\left( \begin{array}{cccc}
0 & \bar{\gamma} (k_x - i k_y) & 0 & 0 \\
\bar{\gamma} (k_x + i k_y) & 0 & 0 & 0 \\
0 & 0 & 0 & \bar{\gamma} (k_x + i k_y) \\
0 & 0 & \bar{\gamma} (k_x - i k_y) & 0 
\end{array} \right),
\eneq
$\mbi{k} = (k_x, k_y)$, and $\Psi_{\mbi{k},\sigma}$ is an annihilation 
operator with four components.  In the operator $\Psi$, the fist and 
second columns indicate an electron at the A and B sublattice 
points around the Fermi point $K$ of the graphite, respectively.  
The third and fourth columns are an electron at the A and B
sublattices around the Fermi point $K'$.  The quantity $\bar{\gamma}$
is defined as $\bar{\gamma} \equiv (\sqrt{3}/2) a \gamma_0$,
where $a$ is the bond length of the graphite plane and $\gamma_0$
($\simeq$ 2.7 eV) is the resonance integral between neighboring 
carbon atoms.  When the above matrix is diagonalized, we
obtain the dispersion relation $E_\pm = \pm \bar{\gamma}
\sqrt{k_x^2 + \kappa_{\nu \phi}^2 (n)}$, where $k_x$ is parallel
with the axis of the nanotube, $\kappa_{\nu \phi} (n) = (2 \pi / L)
(n + \phi - \nu/3)$, $L$ is the circumference length of the 
nanotube, $n$ ($= 0$, $\pm 1$, $\pm 2$, ...) is the index of bands, 
$\phi$ is the magnetic flux in units of the flux quantum, 
and $\nu$ ($= 0$, 1, or 2) specifies the boundary condition 
in the $y$-direction.  The metallic and semiconducting
nanotubes are characterized by $\nu = 0$ and $\nu = 1$ (or 2),
respectively.  Hereafter, we consider the case
$\phi = 0$ and the metallic nanotubes $\nu = 0$.

The second term in Eq. (1) is the impurity model:
\beeqa
H_{\rm imp} &=& E_d \sum_\sigma d_\sigma^\dagger d_\sigma
+ \frac{U_d}{2} \sum_\sigma n_{d,\sigma} n_{d,-\sigma} \nonumber \\
&+& V \sqrt{\frac{2}{N_s}} \sum_{\mbi{k},\sigma}
[ d_\sigma^\dagger ( e^{i(\mbi{k}+\mbi{K}) \cdot \mbi{r}_0}
\Psi_{\mbi{k},\sigma}^{(1)} 
+ e^{i \eta} e^{i(\mbi{k}+\mbi{K'}) \cdot \mbi{r}_0}
\Psi_{\mbi{k},\sigma}^{(3)}) + {\rm h.c.}],
\eneqa
where $d$ is an operator of the localized electron with
spin $\sigma$; its site energy is $E_d$; $U_d$ is the onsite 
repulsive interaction and $n_{d,\sigma}=d_\sigma^\dagger d_\sigma$;
$V$ is the mixing interaction between the electrons on the
nanotube and the localized orbital; $N_s$ is the total
number of lattice sites; $\eta$ is the angle between
the chiral vector of the present nanotube and the tube axis 
direction of the armchair-type nanotubes; and 
$\Psi_{\mbi{k},\sigma}^{(i)}$ ($i=1 - 4$) is the $i$th 
component of the operator $\Psi_{\mbi{k},\sigma}$, assuming 
that the Anderson impurity is located at the site $\mbi{r}_0$ 
of the A sublattice set.  When the impurity is located at one of 
the B sublattice sites, the second and fourth components
of $\Psi_{\mbi{k},\sigma}$ appear in $H_{\rm imp}$.  
However, the result of the single impurity system is 
the same altogether.

The propagator of the electrons on the nanotube is 
defined in the matrix form:
\beeq
G(\mbi{k},\tau) = - \la T_\tau \Psi_{\mbi{k},\sigma} (\tau)
\Psi_{\mbi{k},\sigma}^\dagger (0) \ra,
\eneq
where $T_\tau$ is the time ordering operator with respect
to the imaginary time $\tau$ and 
$\Psi_{\mbi{k},\sigma} (\tau) = {\rm exp} (H\tau)
\Psi_{\mbi{k},\sigma} {\rm exp} (-H\tau)$.  The Fourier 
transform of $G$ is calculated as:
\beeq
G^{-1} (\mbi{k},i\omega_n) = 
\left( \begin{array}{cc}
G_K^{-1} & 0 \\
0 & G_{K'}^{-1} 
\end{array} \right),
\eneq
where $\omega_n = (2n+1) \pi T$ is the odd Matsubara frequency
for fermions.  The components of $G$ are written explicitly:
\beeq
G_K^{-1} (\mbi{k},i\omega_n) =
\left( \begin{array}{cc}
i\omega_n & -\bar{\gamma} (k_x - i k_y) \\
-\bar{\gamma} (k_x + i k_y) & i\omega_n
\end{array} \right),
\eneq
and
\beeq
G_{K'}^{-1} (\mbi{k},i\omega_n) =
\left( \begin{array}{cc}
i\omega_n & -\bar{\gamma} (k_x + i k_y) \\
-\bar{\gamma} (k_x - i k_y) & i\omega_n
\end{array} \right).
\eneq

There are several theoretical characteristic parameters.  
We will explain them as follows:
\begin{itemize}
\item
The total carbon number $N_s$ is given by
\beeq
N_s = A \times L \div (\frac{\sqrt{3}}{2} a^2 ) \times 2
= \frac{4AL}{\sqrt{3}a^2},
\eneq
where $A$ is the length of the nanotube, and $(\sqrt{3}/2) a^2$
is the area of the unit cell.  There are two carbons
in one unit cell, so the factor 2 is multiplied.

\item
The density of states near the Fermi energy $E=0$ is 
calculated as
\beeq
\rho(E) = \frac{A}{2\pi} \int^\infty_{-\infty}
dk_x \delta(E-\bar{\gamma}k_x)
= \frac{aN_s}{4\pi L \gamma_0}.
\eneq
Because two sites in the discrete model correspond
to one site in the continuum $\kp$ model,
the density of sites in the continuum model is given by:
\beeq
\rho(E) = \frac{a}{2 \pi L \gamma_0}.
\eneq

\item
In the theory of Kondo effect, it is necessary to define
the band cutoff $D$ in order that $\rho(E=0) = 1/2D$.
Thus, the quantity $D$ is written explicitly
\beeq
D = \frac{\pi L}{a} \gamma_0.
\eneq
Table I shows several combinations of parameters,
$D$ and $E_1 \equiv \bar{\gamma} \kappa_{00}^2(n=1) 
= (\sqrt{3}\pi a/L) \gamma_0$, as a reference 
for realistic parameter values of nanotubes.
We find that $D$ is a fairly large cutoff for 
typical single wall nanotubes.  The energy $E_1$ at the
bottom of the conduction band $n=1$ is apparently 
larger than the energy scale which corresponds to
room temperatures or low temperatures.  This would mean
that effects of the upper or lower bands which do not
cross the Fermi energy are small enough to neglect 
their contributions to the multi channel Kondo behavior
in low temperatures.
\end{itemize}

In the resolvent method [27,28], we mainly consider the
limit $U_d \rightarrow \infty$ so that double occupancy
at the localized level is forbidden.  The empty state
and the singly occupied state at the magnetic impurity
are taken into account in the formalism.  The spin 
degeneracy $N$ and the number of scattering channels $M$ 
of the band states are generalized to take an arbitral 
integer number.  In the limit of $M, N, \rightarrow 
\infty$ with $\gamma \equiv M/N$ fixed, the single 
impurity problem is represented by the coupled
integral equation:
\beeq
\Sigma_d (\omega) = \frac{\gamma \tilde{\Gamma}}{\pi}
\int d\epsilon f(\epsilon) \Phi_b (\epsilon+\omega),
\eneq
\beeq
\Pi_b (\omega) = \frac{\tilde{\Gamma}}{\pi}
\int d\epsilon f(\epsilon) \Phi_d (\epsilon+\omega),
\eneq
where $\tilde{\Gamma} \equiv \pi \rho N V^2$,
$f(\epsilon) = 1/[{\rm exp}(\epsilon/T)+1]$,
\beeq
\Phi_d (i\omega_n) = \frac{1}{i\omega_n - E_d - \Sigma_d(i\omega_n)}
\eneq
is the resolvent for the singly occupied $d$-state, and
\beeq
\Phi_b (i\nu_n) = \frac{1}{i\nu_n - \Phi_b(i\nu_n)}
\eneq
is the resolvent for the empty $d$-state.  Here, we
denote the even Matsubara frequency as $\nu_n$.

The validity of the multi channel problem for the application
of metallic nanotube can be understood easily by looking 
at the formula of the self energies.  In the case $U_d = 0$,
the self energy of the $d$-electron is
\beeq
\Sigma_d (i \omega_n) = V^2 \frac{2}{N_s} 
\sum_{\mbi{k}} [ G_K^{(1)} (\mbi{k},i \omega_n) 
+ G_{K'}^{(3)} (\mbi{k},i \omega_n)],
\eneq
where $G_K^{(1)}$ is the first diagonal component of the
propagator Eq. (6) and $G_{K'}^{(3)}$ is the third component.
After inserting their explicit forms, we obtain
\beeqa
\Sigma_d(i\omega_n) &=& V^2 \frac{2}{N_s} \sum_{\mbi{k}}
\frac{2 i\omega_n}{(i\omega_n)^2 - \bar{\gamma}^2
(k_x^2 + k_y^2)} \nonumber \\
&=& V^2 \frac{2}{N_s} \sum_{\mbi{k}}
( \frac{1}{i\omega_n - \bar{\gamma}\sqrt{k_x^2 + k_y^2}}
+ \frac{1}{i\omega_n + \bar{\gamma}\sqrt{k_x^2 + k_y^2}}) \nonumber \\
&\simeq& M V^2 \frac{2}{N_s} \sum_{\mbi{k}}
\frac{1}{i\omega_n - \epsilon_{\mbi{k}}},
\eneqa
where the scattering channel number is expressed as $M$ 
and $-D < \epsilon_{\mbi{k}} < D$ with the density 
of states $\rho = 1/2D$ ($D$ is defined by Eq. (12)).
Only the two bands which cross the Fermi energy are
retained in the sum of the last line, because the 
characteristic energy $E_1$ is fairly large as discussed above.
Similarly, in the limit $U_d \rightarrow \infty$, the self 
energy of the resolvent $\Phi_d$ is
\beeq
\Sigma_d(i\omega_n) = - MV^2 \frac{2}{N_s} \sum_{\mbi{k}}
T \sum_{\omega_{n'}} \frac{1}{i\omega_{n'} - \epsilon_{\mbi{k}}}
\Phi_b (i\omega_n - i\omega_{n'})
\eneq
The difference from the $U_d = 0$ case is that the
resolvent $\Phi_b$ appears in the right hand side.
After taking the frequency sum, we obtain the integral
form Eq. (13).  Therefore, we have shown that the
set of integral equations of the multi channel Kondo
problem is valid for an magnetic impurity in metallic
carbon nanotubes.

\section{Low temperature solution of resolvents}

The set of integral equation Eqs. (13,14) can be solved
analytically in the limit of low frequency and low
temperatures [27,28].  Before discussing physical quantities,
we derive analytic forms of resolvents.  As the resolvents
mean singly occupied states and empty state physically,
the analytic property is not the same as that of the
usual propagators.  We define spectral functions for
the empty states at $\omega > E_0$,
\beeq
A_{d,b}^{(+)} (\omega) \theta(\omega-E_0) \equiv
- \frac{1}{\pi} {\rm Im} \Phi_{d,b} (\omega + i\delta),
\eneq
and spectral functions for the occupied states
at $\omega < E_0$,
\beeq
A_{d,b}^{(-)} \equiv {\rm lim}_{T \rightarrow 0}
[ - \frac{1}{\pi} {\rm Im} \Phi_{d,b} (\omega + i\delta) ]
{\rm exp} [\beta(E_0 - \omega)],
\eneq
where $E_0$ is the ground state energy of the magnetic
impurity at zero temperature, $\delta$ is positive
infinitesimal, and $\beta = 1/T$.

After some calculations following Ref. [27], we find
the following formula at low frequency:
\beeq
A_d^{(\pm)} \sim | \Theta (\omega) |^{-\gamma}
\eneq
and
\beeq
A_b^{(\pm)} \sim | \Theta (\omega) |^{-1}.
\eneq
Here,
\beeq
\Theta (\omega) \equiv [ (\frac{1 + \gamma}{\gamma})
(\frac{E_0 - \omega}{T_K})]^\frac{1}{1+\gamma}
\eneq
with $\gamma = M/N$, and $T_K$ is the Kondo temperature:
\beeq
T_K = D (\frac{\gamma \tilde{\Gamma}}{\pi D})^\gamma
{\rm exp} (\frac{\pi E_d}{\tilde{\Gamma}}).
\eneq

For the special case of interests of metallic nanotubes, 
$M=2$ and $N=2$.  Therefore, we find the singular 
frequency dependence around the ground state energy $E_0$:
\beeq
A_d^{(\pm)} (\omega) \sim A_b^{(\pm)} (\omega)
\sim | E_0 - \omega |^\frac{-1}{2}.
\eneq
This singular dependence is the main conclusion of 
this section.  The frequency dependence is shown
schematically in Fig. 1.

\section{Density of states}

The local density of states of $d$-electron is calculated
by the convolution of spectral functions $A_{d,b}^{(\pm)} (\omega)$
[27,28].  The explicit form of the density of states for spin 
$\sigma$ and the channel $\alpha$ at $T=0$ becomes
\beeq
\rho_{\sigma,\alpha} (\omega,0) \simeq 
[ \frac{\pi}{(1+\gamma)^2} N \tilde{\Gamma}]
[1 + \theta(\omega) f_+ (\tilde{\omega})
+ \theta(-\omega) f_- (\tilde{\omega}) ],
\eneq
where
\beeq
f_\pm (\tilde{\omega}) = a_\pm |\tilde{\omega}|^{\Delta_{\rm sp}}
+ b_\pm |\tilde{\omega}|^{\Delta_{\rm ch}},
\eneq
with
\beeqa
a_- &=& - (\frac{4\gamma}{2 + \gamma}\pi)
{\rm sin}(\pi \Delta_{\rm ch}) B(2\Delta_{\rm sp},\Delta_{\rm ch}),\\
a_+ &=& - {\rm cos} (\pi \Delta_{\rm ch}) a_-,\\
b_+ &=& - (\frac{4 W_{\rm ch}}{1 + 2\gamma}\pi)
{\rm sin}(\pi \Delta_{\rm ch}) B(2\Delta_{\rm ch},\Delta_{\rm sp}),\\
b_- &=& {\rm cos} (\pi \Delta_{\rm ch}) b_+.
\eneqa
Here, $\tilde{\omega} \equiv [(1+\gamma)/\gamma](\omega/T_K)$,
$B(x,y)$ is the Beta function, and $W_{\rm ch} \equiv 
\pi T_K/\tilde{\Gamma}$ is the weight of channel fluctuations.
Further, $\Delta_{\rm sp} \equiv 1/(1+\gamma)$ and 
$\Delta_{\rm ch} \equiv \gamma/(1+\gamma)$ are the
scaling dimensions of spin and channel fields, respectively.
As $\Delta_{\rm ch} \propto M$, this measures the magnitude
of fluctuations from the channel degree of freedom.
Also, $\Delta_{\rm sp} \propto N$ means that this is
a measure of contributions from the spin degree of freedom.
Both quantities determine the degree of singularities of
electronic density of states and physical quantities at
low frequencies.  They are the most important parameters
introduced in this section.

Specially for metallic carbon nanotubes, we know that
$\Delta_{\rm sp} = \Delta_{\rm ch} = 1/2$.  This implies
the singularity around the Fermi energy $\omega = 0$:
\beeqa
\rho (\omega,0) &\sim& 1 + \theta(\omega) |\omega|^{\frac{1}{2}}
+ \theta(-\omega) |\omega|^{\frac{1}{2}}, \nonumber \\
&\sim& \sqrt{|\omega|}.
\eneqa
Such the singular functional form implies that a pseudo gap
develops at the top of the Kondo resonance peak which appear
at temperatures much lower than $T_K$.  There appears a dip 
in the density of states at the Fermi energy.  The dip
structure of the density of states is shown schematically
in Fig. 2.  This is the local non-Fermi liquid behavior 
discussed in detail in the literatures [24,29].  If it is 
possible to measure the local density of states of a metallic 
atom attached to the carbon nanotubes, for example, by 
scanning tunneling microscope, we could observe such the 
pseudo gap behavior when role of the multi channel 
scatterings is dominant.

\section{Resistivity}

In this section, we consider the electric resistivity in 
order to look at how the singular behavior will be observed.
The scattering rate $\tau$ is calculated from the scattering
$t$ matrix:
\beeqa
\tau_{\sigma,\alpha}(\omega,T)^{-1} &=&
-2 {\rm Im} t_{\sigma,\alpha}^{(1)}(\omega+i\delta,T), \nonumber \\
&=& \frac{2 \tilde{\Gamma} \rho_{\sigma,\alpha}(\omega,T)}{\rho N}.
\eneqa
The relation with the electronic resistivity
\beeq
\bar{\rho}(T) \sim [ \int d\epsilon 
( - \frac{\partial f}{\partial \epsilon})
\tau(\epsilon,T)]^{-1}
\eneq
gives the low temperature behavior:
\beeq
\frac{\bar{\rho}(T)}{\bar{\rho}(0)} \sim
1 - c (\frac{T}{T_K})^{{\rm min} (\Delta_{\rm sp},\Delta_{\rm ch})}
+ ... ,
\eneq
where $c$ is a constant, but it is difficult to obtain
its explicit form only from the information of $\omega$-dependence
of $\rho_{\sigma,\alpha}$.  Here, we stress that the above singular 
functional form agrees with that obtained from the conformal field 
approach [30,31].  Therefore, the temperature dependence of the
leading term $(T/T_K)^{{\rm min} (\Delta_{\rm sp},\Delta_{\rm ch})}$
is a general result which is independent from method of theoretical 
treatment.

For the metallic carbon nanotubes, we already know 
$\Delta_{\rm sp} = \Delta_{\rm ch} = 1/2$.  Therefore, 
the low temperature behavior
\beeq
\frac{\bar{\rho}(T)}{\bar{\rho}(0)} \sim
1 - c \sqrt{\frac{T}{T_K}}
\eneq
is expected from the above general formula.

\section{Magnetic susceptibility}

Spin and channel susceptibilities are calculated by the
linear response function.  The spin susceptibility is
the magnetic susceptibility in other words.  The imaginary
parts of dynamical susceptibilities are defined as
\beeq
\tilde{\chi}''_{\rm sp} = \frac{1}{N} {\rm Im} \chi_{\rm sp}
\eneq
and
\beeq
\tilde{\chi}''_{\rm ch} = \frac{1}{M} {\rm Im} \chi_{\rm ch}.
\eneq

The first term of $\tilde{\chi}''_{\rm sp}$ at $T=0$ is 
calculated to be
\beeq
\tilde{\chi}''_{\rm sp} (\omega,0) \sim
\frac{C_{\rm sp}}{T_K} {\rm sgn} \omega 
|\tilde{\omega}|^{(\Delta_{\rm sp}-\Delta_{\rm ch})},
\eneq
where
\beeq
C_{\rm sp} = \gamma \Delta_{\rm sp}^2 {\rm sin} (\pi \Delta_{\rm sp})
B(\Delta_{\rm sp},\Delta_{\rm sp}).
\eneq
The second correction gives the $\omega$ dependence:
\beeq
\tilde{\chi}''_{\rm sp} (\omega,0) 
\sim |\tilde{\omega}|^{(2\Delta_{\rm sp}-\Delta_{\rm ch})}.
\eneq

In the similar way, the dominant term of 
$\tilde{\chi}''_{\rm ch}$ at $T=0$ becomes
\beeq
\tilde{\chi}''_{\rm ch} (\omega,0) \sim
\frac{C_{\rm ch}}{T_K} {\rm sgn} \omega 
|\tilde{\omega}|^{(\Delta_{\rm ch}-\Delta_{\rm sp})},
\eneq
where
\beeq
C_{\rm ch} = W_{\rm ch}^2 \Delta_{\rm sp}
{\rm sin} (\pi \Delta_{\rm ch})
B(\Delta_{\rm ch},\Delta_{\rm ch}).
\eneq
The second term has the $\omega$ dependence:
\beeq
\tilde{\chi}''_{\rm ch} (\omega,0) 
\sim |\tilde{\omega}|^{(2\Delta_{\rm ch}-\Delta_{\rm sp})}.
\eneq

The above general formulas Eqs. (40-45) reduce to that of 
metallic carbon nanotubes.  The result of singular behavior is 
common for spin and channel susceptibilities:
\beeq
\tilde{\chi}'' (\omega,0) 
\sim A {\rm sgn} \omega (1 - B \sqrt{\frac{|\tilde{\omega}|}{T_K}}
+ ... ),
\eneq
where $A$ and $B$ are constants.  We find $\sqrt{|\omega|}$
dependence at low frequencies.  We note that the term
sgn $\omega$ is related with the real part:
\beeq
\tilde{\chi}' (\omega,T) \sim
- {\rm log} [ \frac{{\rm max} (\omega,T)}{T_K} ].
\eneq

\section{Discussion}

We have discussed the solutions of the integral equations
in the limit $M, N \rightarrow \infty$ in this paper.
On the other hand, the higher order corrections of O($1/N$) 
[27,28] are known to change the coefficients of the 
leading terms of singular behaviors, but they do not 
affect the power of singularities.  Therefore, the 
non-Fermi liquid behaviors are not affected by the
higher order corrections by the O($1/N$) expansion.

In the course of the investigation, we have neglected
the valence and conduction band states which do not
cross the Fermi energy.  This approximation applies
well at low energies $|\omega| \sim T_K \ll E_1$.
Typical magnitudes of $E_1$ have been shown in Table I.
However, at higher energies or at high temperatures,
such the deep bands will give some corrections to
low energy singular behaviors.  Furthermore, functional
forms of spectral functions at entire frequencies
can be affected by the presence of deep bands.
In order to treat the deep band effects, we have
to solve the resolvent equations numerically at least,
even though it is beyond the scope of this paper.

There are two scattering channels when the impurity
interacts with the single wall metallic nanotube.
Is it possible that there are more scattering channels?
When two metallic carbon nanotubes are present and
one impurity interacts with both nanotubes, four
scattering channels are realized as long as the 
interactions between electrons of two nanotubes are 
negligible.  The powers of the singularities of
the density of states and physical quantities become
different from those of the case of the single nanotube.
By putting $M=4$ and $N=2$, the general formulas of
this paper can be applied to one magnetic impurity
which interacts with two aligned metallic nanotubes.
If the impurity interacts with more nanotubes,
such the general formulas are useful in order to
predict singular behaviors which depend on the 
scattering channel number.

In the literature, for example in [32], the power law
temperature dependences of electronic resistivity have
been interpreted by the Luttinger liquid picture
of the correlated one-dimensional systems.  The Kondo
effect in the Luttinger liquid has been studied
by Furusaki and Nagaosa [33].  A further study of the 
Kondo impurity with two scattering channels in the 
presence of conduction bands with Luttinger liquid 
properties is a fascinating extention of the 
present work.

\section{Summary}

Magnetic impurity effects on metallic carbon nanotubes
have been investigated theoretically.  The resolvent 
method for the multi channel Kondo effect has been applied 
to the band structure of the $\kp$ model in the limit of the
infinite onsite repulsion at the impurity site.  We have
discussed the local non-Fermi liquid behavior which could
be observed at temperatures lower than the Kondo temperature 
$T_{\rm K}$.  The density of states of localized electron 
has a singularity $\sim |\omega|^{1/2}$.  This singular
behavior gives rise to a pseudo gap at the Kondo resonance
in low temperatures.  The temperature dependence of the 
electronic resistivity is predicted as $T^{1/2}$, and the
imaginary part of dynamical susceptibilities has the
$|\omega|^{1/2}$ dependence.

\mbox{}

\begin{flushleft}
{\bf Acknowledgements}
\end{flushleft}

\noindent
Useful discussion with the members of Condensed Matter
Theory Group\\
(\verb+http://www.etl.go.jp/+\~{}\verb+theory/+),
Electrotechnical Laboratory is acknowledged.

\pagebreak
\begin{flushleft}
{\bf References}
\end{flushleft}

\noindent
$[1]$ M. S. Dresselhaus, G. Dresselhaus, and P. C. Eklund,
``Science of Fullerenes and Carbon Nanotubes",
(Academic Press, San Diego, 1996).\\
$[2]$ R. Saito, G. Dresselhaus, and M. S. Dresselhaus,
``Physical Properties of Carbon Nanotubes",
(Imperial College Press. London, 1998).\\
$[3]$ J. W. Mintmire, B. I. Dunlap, and C. T. White,
Phys. Rev. Lett. {\bf 68}, 631 (1992).\\
$[4]$ N. Hamada, S. Sawada, and A. Oshiyama,
Phys. Rev. Lett. {\bf 68}, 1579 (1992).\\
$[5]$ R. Saito, M. Fujita, G. Dresselhaus, and M. S. Dresselhaus,
Appl. Phys. Lett. {\bf 60}, 2204 (1992).\\
$[6]$ K. Tanaka, K. Okahara, M. Okada, and T. Yamabe,
Chem. Phys. Lett. {\bf 193}, 101 (1992).\\
$[7]$ K. Harigaya, Phys. Rev. B {\bf 45}, 12071 (1992).\\
$[8]$ K. Harigaya and M. Fujita, Phys. Rev. B {\bf 47},
16563 (1993).\\
$[9]$ J. W. G. Wild\"{o}er, L. C. Venema, A. G. Rinzler,
R. E. Smalley, and C Dekker. Nature {\bf 391}, 59 (1998).\\
$[10]$ T. W. Odom, J. L. Huang, P. Kim, and C. M. Lieber,
Nature {\bf 391}, 62 (1998).\\
$[11]$ P. Avouris et al, in ``Electronic Properties of Novel Materials:
Science and Technology of Molecular Nanostructures",
ed. H. Kuzmany, (AIP, New York, 1999), p. 393.\\
$[12]$ G. Bergmann, Phys. Rep. {\bf 1}, 1 (1984).\\
$[13]$ Y. Shibayama et al, Mol. Cryst. Liq. Cryst. {\bf 310},
273 (1998).\\
$[14]$ K. Takai et al, Mol. Cryst. Liq. Cryst. (to be published).\\
$[15]$ R. Saito, M. Yagi, T. Kimura, G. Dresselhaus, and
M. S. Dresselhaus, J. Phys. Chem. Sol. {\bf 60}, 715 (1999).\\
$[16]$ W. A. Phillips, J. Low Temp. Phys. {\bf 7}, 161 (1971).\\
$[17]$ P. W. Anderson, B. I Halperin, and C. M. Varma,
Philos. Mag. {\bf 25}, 1 (1972).\\
$[18]$ J. Kondo, Physica (Amsterdam) {\bf 84B}, 207 (1976).\\
$[19]$ J. L. Black and B. L. Gyroffy. Phys. Rev. Lett. {\bf 41},
1595 (1978).\\
$[20]$ K. Vladar and A. Zawadowski, Phys. Rev. B {\bf 28}, 
1564 (1983); ibid {\bf 28}, 1956 (1983).\\
$[21]$ K. Harigaya, cond-mat/9810341.\\
$[22]$ K. Harigaya, Phys. Rev. B {\bf 60}, 1452 (1999).\\
$[23]$ The slave boson method has been used for the heavy 
fermion ($f$-electron) systems in K. Harigaya, 
J. Phys.: Condens. Matter {\bf 2}, 3259 (1990); 
K. Harigaya, ibid {\bf 2}, 4623 (1990).\\
$[24]$ L. Degiorgi, Rev. Mod. Phys. {\bf 71}, 687 (1999).\\
$[25]$ H. Ajiki and T. Ando, J. Phys. Soc. Jpn. {\bf 62},
1255 (1993).\\
$[26]$ T. Ando and T. Nakanishi, J. Phys. Soc. Jpn. {\bf 67},
1704 (1998).\\
$[27]$ E. M\"{u}ller-Hartmann, Z. Phys. B {\bf 57}, 281 (1984).\\
$[28]$ P. Coleman, Phys. Rev. B {\bf 29}, 3035 (1984).\\
$[29]$ D. L. Cox and A. Zawadowski, Adv. in Phys. {\bf 47}, 599 (1998).\\
$[30]$ I. Affleck and A. W. W. Ludwig, Nucl. Phys. B {\bf 352},
849 (1991); ibid {\bf 360}, 641 (1991).\\
$[31]$ A. W. W. Ludwig and I. Affleck, Phys. Rev. Lett. {\bf 67}, 
3160 (1991).\\
$[32]$ M. Bockrath et al, Nature {\bf 397}, 598 (1999).\\
$[33]$ A. Furusaki and N. Nagaosa, Phys. Rev. Lett.
{\bf 72}, 892 (1994).\\

\pagebreak

\noindent
TABLE I.  Parameters $D$ and $E_1$ for typical 
metallic carbon nanotubes.\\
\\
\begin{tabular}{ccc} \hline \hline
Nanotube & $D$ ($\gamma_0 =$ 2.7 eV) & $E_1$ \\ \hline
(5,5)    & 8.60 $\gamma_0 =$ 23.2 eV & 1.99 $\gamma_0 =$ 5.37 eV \\
(10,10)  & 12.2 $\gamma_0 =$ 32.9 eV & 1.40 $\gamma_0 =$ 3.78 eV \\
(15,15)  & 14.9 $\gamma_0 =$ 40.2 eV & 1.15 $\gamma_0 =$ 3.11 eV \\ \hline
(6,0)    & 16.3 $\gamma_0 =$ 44.0 eV & 1.05 $\gamma_0 =$ 2.84 eV \\
(9,0)    & 24.5 $\gamma_0 =$ 66.2 eV & 0.698 $\gamma_0 =$ 1.88 eV \\
\hline \hline
\end{tabular}

\pagebreak

\noindent
\begin{flushleft}
{\bf Figure Captions}
\end{flushleft}

\mbox{}

\noindent
Fig. 1.  The singular behavior of the spectral functions 
$A_{d,b}^{(\pm)} (\omega)$.  The figure is drawn schematically.

\mbox{}

\noindent
Fig. 2.  The dip structure of the density of states 
$\rho(\omega)$.  The figure is drawn schematically.

\end{document}